\documentclass[aip,amsmath,reprint,amssymb]{revtex4-1}

\usepackage{graphicx}
\usepackage{dcolumn}
\usepackage{bm}
\usepackage{color}

\usepackage[utf8]{inputenc}
\usepackage[T1]{fontenc}
\usepackage{mathptmx}
\usepackage{etoolbox}

\makeatletter
\def\@email#1#2{%
 \endgroup
 \patchcmd{\titleblock@produce}
  {\frontmatter@RRAPformat}
  {\frontmatter@RRAPformat{\produce@RRAP{*#1\href{mailto:#2}{#2}}}\frontmatter@RRAPformat}
  {}{}
}%
\makeatother

\begin{document}


\title{Theory of electric reactance emerging from spin Hall effect} 



\author{Yasufumi Araki}
\email{araki.yasufumi@jaea.go.jp}
\affiliation{Advanced Science Research Center, Japan Atomic Energy Agency, Tokai 319-1195, Japan}

\author{Jun'ichi Ieda}
\affiliation{Advanced Science Research Center, Japan Atomic Energy Agency, Tokai 319-1195, Japan}


\date{\today}

\begin{abstract}
The spin Hall effect in a heavy metal intercorrelates an AC electric current to the magnetization dynamics in an adjacent ferromagnet,
which manifests as an electric reactance in the system's current-voltage response.
We present a comprehensive theoretical analysis for this emergent reactance contribution
in the frequency regime relevant to transport measurements up to a few GHz.
Our analysis reveals that the reactance becomes inductor-like at low frequency below the ferromagnetic resonance.
Crucially, we find that the sign of the reactance is directly governed by the spin transfer mechanism at the interface,
which depends on the competition between its damping-like and field-like components parametrized by the spin mixing conductance.
This characteristic behavior in the reactance offers a powerful transport observable in distinguishing the interfacial spin transfer processes in spintronic materials.
\end{abstract}


\maketitle 



%
%

%



Reactance is the electric response inducing a phase shift between an AC current $I_\omega$ and voltage $V_\omega$
depending on the frequency $\omega$.
It is given by the imaginary part ${\rm Im} \ Z_\omega$ in the complex impedance $Z_\omega = V_\omega/I_\omega$.
While the resistance ${\rm Re} \ Z_\omega$ leads to the energy dissipation due to the Joule heating,
the reactance exerts no work and conserves the energy of the circuit averaged over a cycle.
In electronic circuits in devices, reactance is commonly implemented by inductors and capacitors.
Inductor with its inductance $L$ gives $Z_\omega = i\omega L$ based on the Faraday's electromotive force from a magnetic flux inside a solenoid,
whereas capacitor with its capacitance $C$ gives $Z_\omega = 1/i\omega C$ based on the displacement current from a polarization in a dielectric medium.

While the conventional inductors and capacitors rely on classical dynamics of electromagnetic fields,
recent studies have discovered the quantum contribution to a reactance emerging from spin dynamics in magnetic materials \cite{nagaosa2019emergent,yokouchi2020emergent,kitaori2021emergent,kurebayashi2021electromagnetic,ieda2021intrinsic,yamane2022theory,araki2023emergence,furuta2023symmetry,kitaori2023doping,yokouchi2023emergent,kitaori2024enhanced,matsushima2024emergent,oh2024emergent,anan2025emergent}.
By interconversion processes between charge and spin dynamics of electrons,
an AC current or voltage is converted into the spin dynamics and vice versa,
which contributes to an AC current-voltage response.
Focusing on its low-frequency characteristics,
such an effect is often termed as the ``emergent inductance''.
In metallic magnets with nonuniform textures such as magnetic spirals or domain walls,
the dynamics of magnetic texture excited by the current-induced spin-transfer torque (STT) \cite{slonczewski1996current,berger1996emission} yields
a voltage response from the spinmotive force (SMF) \cite{volovik1987linear,barnes2007generalization,nagaosa2012emergent},
and hence gives a reactance \cite{nagaosa2019emergent,yokouchi2020emergent,kitaori2021emergent,kurebayashi2021electromagnetic,ieda2021intrinsic,furuta2023symmetry,kitaori2023doping,kitaori2024enhanced,matsushima2024emergent,anan2025emergent}.
Even for a uniform ferromagnet (FM) with no textures,
a reactance can emerge from a spin-orbit coupling (SOC),
via the current-induced spin-orbit torque (SOT) \cite{manchon2008theory,manchon2009theory,mihai2010current} and the SOC-oriented SMF \cite{kim2012prediction,yamane2013spinmotive,ciccarelli2015magnonic}.
Such an effect is expected in noncentrosymmetric electronic systems,
typically the interfaces between magnets and nonmagnetic materials like heavy metals or semiconductors \cite{yamane2022theory},
and also the edge states of topological insulators \cite{araki2023emergence}.
Other types of spin dynamics,
such as dynamics of antiferromagnetic orders \cite{yokouchi2023emergent} and spin fluctuations in paramagnetic states \cite{oh2024emergent},
can also contribute to the emergence of reactance.
\textcolor{black}{
    With such a growing interest on reactance,
    it is necessary to know possible mechanisms contributing to reactance,
    to identify the origin of experimentally measured signals.
}

\textcolor{black}{
    As a typical spin-charge interconversion effect that may participate in the reactance,
    we here focus on possible role of the spin Hall effect (SHE).
}
The SHE is the interconversion of charge and spin currents transverse to each other,
which originates from the SOC in the bulk of heavy metals (HMs), semiconductor quantum wells, and some topological materials \cite{dyakonov1971possibility,hirsch1999spin,murakami2003dissipationless,sinova2004universal,sinova2015spin}.
To manipulate and detect spin currents in devices, the SHE is broadly used in realizing various spintronics functionalities \cite{jungwirth2012spin}.
The SHE results in some electronic transport properties typically in HM/FM heterostructures.
One of the most typical effects among them is the spin Hall magnetoresistance (SMR),
where the DC resistivity gets modulated by the interfacial reflection or absorption of the spin Hall current
depending on the magnetization direction in the FM \cite{weiler2012local,huang2012transport,nakayama2013spin,chen2013theory,chen2016theory,kato2020microscopic,brataas2013insulating}.
The SHE contribution to the optical response in the THz spectroscopy is also studied in recent theory \cite{reiss2021theory} and experiment \cite{kubavsvcik2025observation}.
Besides,
the reactance in the transport frequency regime, ranging from a few Hz up to the orders of GHz,
is also expected convey a great information of the spin-charge conversion processes like the SMF,
as suggested from the preceding theoretical and experimental studies of emergent inductances.
While the transport impedance in the HM/FM heterostructure was experimentally reported \cite{lotze2014spin},
a thorough theoretical quantification of the transport reactance in connection with the SHE-induced magnetization dynamics
is left to be elucidated in the transport regime.

In this work, we theoretically analyze the reactance in the HM/FM heterostructure
emerging from the combination of SHE and magnetization dynamics,
focusing on the transport regime up to the GHz orders.
We investigate its dependence on the macroscopic material parameters,
including the size of the system, magnetic properties,
and the spin mixing conductance characterizing the spin transfer at the HM/FM interface.
Akin to the emergent inductances found in previous studies,
the reactance becomes proportional to the frequency below the ferromagnetic resonance (FMR),
exhibiting an inductor-like behavior.
In particular, the sign of the reactance is governed by the real and imaginary parts of spin mixing conductance,
corresponding to the damping-like and field-like components of spin injection pumping processes.
This sign change behavior can be compared to the discussions on the sign of emergent inductance in the preceding theories \cite{ieda2021intrinsic,yamane2022theory,kurebayashi2021electromagnetic,anan2025emergent}.
Our study implies that the measurement of reactance is helpful in understanding the nature of
\textcolor{black}{
    the interfacial spin transfer processes in heterostructure systems,
    which are highly material-specific and have remained to be elucidated.
}

\begin{figure}[tbp]
    \centering
    \includegraphics[width=8cm]{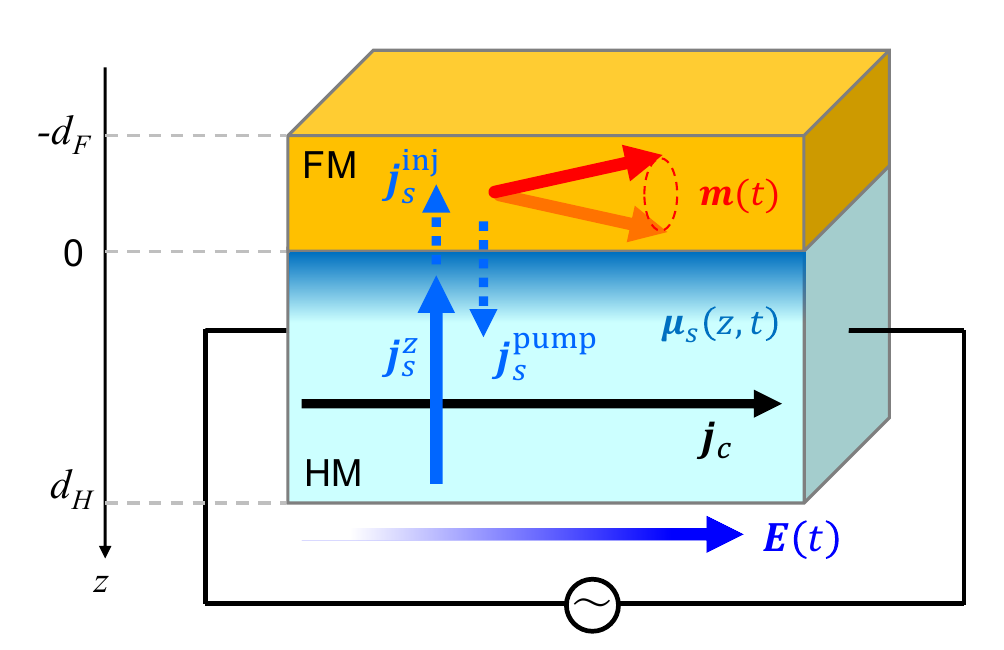}
    \caption{Schematic picture of the hypothetical setup.
    A heterostructure of the films of heavy metal (HM) and insulator of ferromagnet (FM) is taken.
    We consider a linear response between an alternating electric field $\boldsymbol{E}(t)$ and current $\boldsymbol{j}_c(t)$ involving the magnetization dynamics $\boldsymbol{m}(t)$.}
    \label{fig:schematic}
\end{figure}

The algebraic formulation of the reactance can be derived
on the basis of the finite-frequency formulations of the SHE and magnetization dynamics
established in Refs.~\onlinecite{kubavsvcik2025observation} and \onlinecite{chiba2014current}.
We consider a heterostructure of HM and FM films,
as schematically shown in Fig.~\ref{fig:schematic},
where we limit FM to a ferromagnetic insulator that does not conduct a charge current.
We take the Cartesian coordinate $(x,y,z)$
with the $z$-axis taken to the thickness direction.
The thicknesses of the two films are denoted as $d_H$ and $d_F$, respectively,
with their interface at the $z=0$ plane.
We assume the homogeneity of the two films within the film plane,
so that the spatial inhomogeneities of all the physical quantities depend only on $z$.
In this setup,
we take the electric field $\boldsymbol{E}(t) = {\rm Re} [\boldsymbol{E}_\omega e^{i\omega t}]$ alternating with the frequency $\omega$ as an input,
and evaluate the charge current response $\boldsymbol{j}_c(z,t) = {\rm Re} [\boldsymbol{j}_{c\omega}(z) e^{i\omega t}]$
to extract the impedance including the reactance.
\textcolor{black}{
    $\boldsymbol{E}_\omega$ is assumed to be infinitesimal so that the linear response $O(\boldsymbol{E}_\omega)$ should be most dominant in the responses of all the physical quantities considered here.
}


In the HM film,
we take into account the interconversion between the charge and spin currents,
characterized by the dimensionless spin Hall angle $\theta_{\rm SH}$.
The flow of spin current in the $z$-direction, $\boldsymbol{j}_s^z$,
generates the spin accumulation $\boldsymbol{\mu}_s$ inhomogeneous in the $z$-direction.
With the spatially isotropic longitudinal conductivity $\sigma_0 (= \rho_0^{-1})$,
the charge current $\boldsymbol{j}_c =(j_c^x, j_c^y, j_c^z)$,
the spin current $\boldsymbol{j}_s^z =(j_{s_x}^z, j_{s_y}^z, j_{s_z}^z)$,
and the spin accumulation $\boldsymbol{\mu}_s =(\mu_{s_x}, \mu_{s_y}, \mu_{s_z})$,
are governed by the relations,
\begin{align}
    \boldsymbol{j}_{c}(z,t) &= \sigma_0 \boldsymbol{E}(t) + \frac{\theta_{\mathrm{SH}} \sigma_0}{2e} \hat{\boldsymbol{z}}\times\partial_z\boldsymbol{\mu}_{s}(z,t) \label{eq:response-jc} \\
    \boldsymbol{j}^z_{s}(z,t) &= -\frac{\sigma_0}{2e}\partial_z \boldsymbol{\mu}_{s}(z,t) -\theta_{\mathrm{SH}} \sigma_0 \hat{\boldsymbol{z}}\times \boldsymbol{E}(t), \label{eq:response-js}
\end{align}
where the dimensions of $\boldsymbol{j}_{s}^z$ and $\boldsymbol{\mu}_{s}$ are defined to be same as those of $\boldsymbol{j}_{c}$ and electrochemical potential, respectively.
The inhomogeneity of $\boldsymbol{\mu}_{s}(z,t)$ is governed by the continuity equation,
\begin{align}
    \left[\partial_t + {\tau_s^{-1}} \right] \boldsymbol{\mu}_{s}(z,t) &= -\frac{2}{e \nu_e}\partial_z\boldsymbol{j}^z_{s}(z,t) . \label{eq:spin-diffusion}
\end{align}
where $\tau_s$ is the spin relaxation time, and $\nu_e$ is the electron density of states in the HM.
By using Eq.~(\ref{eq:response-js}),
the right-hand side reduces to the spin diffusion term $D \partial_z^2 \boldsymbol{\mu}_s$,
with the diffusion coeffcient $D = \sigma / e^2 \nu_e$.
The spatial inhomogeneity of $\boldsymbol{\mu}_{s}$ from the spin diffusion is characterized by the spin diffusion length, $\lambda_s = \sqrt{D \tau_s}$.
\textcolor{black}{
    Note that the spin diffusion equation cannot describe the processes at microscopic length scales shorter than $\lambda_s$,
    especially the interfacial spin relaxation processes.
    Nevertheless, such effects can be effectively incorporated in the boundary conditions for the spin diffusion equation,
    in terms of the material parameters such as the spin mixing conductance,
    as we shall discuss later.
}

In the FM film, the dynamics of spins in response to the interfacial spin transfer should be considered.
Along with the preceding theories of emergent inductance \cite{yamane2022theory,araki2023emergence},
we assume that the spins in the FM precess coherently,
so that their dynamics is described by a single vector variable $\boldsymbol{m}(t) = \boldsymbol{M}(t)/M_s$.
Here, $\boldsymbol{M}(t)$ is the time-dependent magnetization vector, and $M_s$ is the magnitude of saturation magnetization.
This macrospin approximation is justified
\textcolor{black}{
    for $d_F$ shorter than the magnon wavelength, typically below $10 \mathchar`- 20 \;{\rm nm}$.
    We also require the temperature to be well below
}
the magnon gap so that the contribution from incoherent thermal magnons \cite{reiss2021theory} is negligible.
Under \textcolor{black}{these assumptions}, the spin dynamics in the FM
is governed by the Landau-Lifshitz-Gilbert (LLG) equation,
\begin{align}
    \partial_t\boldsymbol{m} &= -\gamma \boldsymbol{m} \times \boldsymbol{B}_{\mathrm{eff}} + \alpha\boldsymbol{m}\times\partial_t\boldsymbol{m} +\boldsymbol{\tau}^{\rm int}, \label{eq:LLG}
\end{align}
with the gyromagnetic ratio $\gamma$ and the Gilbert damping constant $\alpha$.
The effective field $\boldsymbol{B}_{\mathrm{eff}}$ includes the external magnetic field and the magnetic anisotropy,
while we omit the effect of the Oersted field.
The last term $\boldsymbol{\tau}^{\rm int}$ accounts for the spin-transfer torque (STT) from the interfacial spin current $\boldsymbol{j}_s^{\rm int} \equiv \boldsymbol{j}_s^{z}(z=0)$,
\begin{align}
    \boldsymbol{\tau}^{\rm int}(t) \equiv {\rm Re}\left[\boldsymbol{\tau}^{\rm int}_\omega e^{i\omega t} \right] = \frac{-\gamma}{M_s d_F} \frac{\hbar}{2e} \boldsymbol{m} \times \left[ \boldsymbol{m} \times \boldsymbol{j}_s^{\rm int}(t) \right], \label{eq:t-int}
\end{align}
with $M_s$ the saturation magnetization.
\textcolor{black}{
    Within the linear response to $\boldsymbol{E}_\omega$,
    the magnetization dynamics is also linearized around its equilibrium position $\boldsymbol{m}_0$ as $\boldsymbol{m}(t) = \boldsymbol{m}_0 + {\rm Re}[\boldsymbol{u}_\omega e^{i\omega t}]$,
    and its response to $\boldsymbol{\tau}^{\rm int}_\omega$
}
is symbolically given as
$\boldsymbol{u}_\omega = D_\omega \boldsymbol{\tau}^{\rm int}_\omega$.
\textcolor{black}{
    Here, $D_\omega$ is the Green's function for the magnetization dynamics,
    which 
}
shows a resonance structure around the frequency $\Omega_0$ of the ferromagnetic resonance (FMR).
\textcolor{black}{
    Although the magnetization dynamics under a finite $\boldsymbol{E}_\omega$ may develop nonlinearity near the FMR
    and may alter the resonance spectrum,
    we here mainly focus on the off-resonant region $\omega < \Omega_0$ and will not consider the details of nonlinearity. 
}

The above equations are uniquely solved by setting the boundary conditions.
At the vacuum end of the HM film,
the spin current vanishes, i.e.,
$\boldsymbol{j}_{s}^z(z=d_H) = 0$.
On the other hand, at the interface of the HM and FM films,
the interfacial spin transfer
\textcolor{black}{
    correlates the spin current $\boldsymbol{j}_{s}^z(z=0) = \boldsymbol{j}_s^{\rm int}$ and the magnetization dynamics.
    As often done in the spin-current-based analyses of experimentally measured data,
    we take the phenomenological description on $\boldsymbol{j}_s^{\rm int}$,
    without considering details of microscopic processes at the interface.
    Here, the spin injection and pumping components,
    $\boldsymbol{j}_s^{\rm int} = \boldsymbol{j}_{s}^{\mathrm{inj}} + \boldsymbol{j}_{s}^{\mathrm{pump}}$,
    are formulated as \cite{brataas2000finite},
}
\begin{align}
    \boldsymbol{j}_{s}^{\mathrm{inj}}(t) &= g_{\rm r} \boldsymbol{m}\times ( \boldsymbol{m} \times \boldsymbol{\mu}_{s}^{\rm int} ) + g_{\rm i} \boldsymbol{m} \times \boldsymbol{\mu}_{s}^{\rm int} \label{eq:j-inj} \\
    \boldsymbol{j}_s^{\mathrm{pump}}(t) &= \hbar g_{\rm r} \boldsymbol{m} \times \partial_t \boldsymbol{m}  + \hbar g_{\rm i} \partial_t\boldsymbol{m}, \label{eq:j-pump}
\end{align}
with the interfacial spin accumulation $\boldsymbol{\mu}_s(z=0) = \boldsymbol{\mu}_s^{\rm int}$.
\textcolor{black}{
    The spin mixing conductance $g_{\uparrow\downarrow} = g_{\rm r} + i g_{\rm i}$ parametrizes the spin transfer process specific to the interface,
}
where $g_{\rm r}$ and $g_{\rm i}$ yield the damping-like (adiabatic) and field-like (nonadiabatic) components of the spin injection and pumping, respectively.
\textcolor{black}{
    Note that the transferred spin may get partially relaxed at the interface,
    known as the interfacial spin memory loss,
    whose effect on the macroscopic spin transport is equivalent to the renormalization of $g_{\uparrow\downarrow}$ \cite{rojas2014spin}.
    In the present formulation,
    we take $g_{\uparrow\downarrow}$ as the renormalized one incorporating such an effect of spin memory loss.
}

By solving the above equations at linear order,
we obain $\boldsymbol{j}_{c\omega}(z)$ in response to $\boldsymbol{E}_\omega$
(see Supplemental Information for the detailed derivation process).
With the spatially averaged $\boldsymbol{j}_{c\omega}^{\rm ave} = \tfrac{1}{d_H} \int dz \ \boldsymbol{j}_{c\omega}(z)$,
the impedance per volume (i.e., AC resistivity) of the HM is given
as a frequency-dependent tensor $\rho_\omega$ that satisfies the relation $\boldsymbol{E}_\omega = \rho_{\omega} \boldsymbol{j}_{c\omega}^{\rm ave}$.
The SHE-induced correction to the impedance,
$\Delta\rho_\omega \equiv \rho_\omega - \rho_0$, is written in the tensor form as,
\begin{align}
    \frac{\Delta\rho_\omega}{\rho_0} &= -\theta_{\rm SH}^2 \frac{\lambda_\omega}{d_H} R_{\hat{\boldsymbol{z}}} \biggl[ 2\eta_\omega(\tfrac{d_H}{2}) + \frac{\eta_\omega^2 (\tfrac{d_H}{2}) G_\omega}{\tfrac{\sigma_0}{2e\lambda_\omega} + G_\omega / \eta_\omega(d_H)}  \biggr] R_{\hat{\boldsymbol{z}}}, \label{eq:general-SMI}
\end{align}
with $\lambda_\omega = \lambda_s / \sqrt{1+i\omega\tau_s}$ and $\eta_\omega(x) = \tanh({x}/{\lambda_\omega})$.
We here emply the tensor notation $R_{\boldsymbol{n}}$ as a $90^\circ$-rotation around a unit vector $\boldsymbol{n}$,
i.e., $R_{\boldsymbol{n}} \boldsymbol{v} \equiv \boldsymbol{n}\times\boldsymbol{v}$.
$G_\omega$ is a tensor defined as
\begin{align}
    G_\omega &= \left[ 1 - i\omega \frac{\hbar^2}{2e}\frac{\gamma}{M_s d_F} G_s R_{\boldsymbol{m}_0} D_\omega \right]^{-1} G_s, \label{eq:G-omega}
\end{align}
where $G_s \equiv -g_{\rm r} R_{\boldsymbol{m}_0}^2 -g_{\rm i} R_{\boldsymbol{m}_0}$ is introduced as the tensor form of spin mixing conductance.
We may regard $G_\omega$ as the interfacial spin mixing conductance ``dressed'' with the magnetization dynamics characterized by $D_\omega$.
Indeed, $G_\omega$ reduces to $G_s$ in the DC limit $\omega =0$,
where Eq.~(\ref{eq:general-SMI}) reproduces the form of SMR for the HM/FM heterostructure \cite{chen2013theory}.

The $\omega$-dependence in $\lambda_\omega$ and $G_\omega$ makes the impedance complex and generates a finite reactance.
As long as the transport measurement up to GHz regime is considered,
the $\omega$-dependence in $\lambda_\omega$ is negligible,
because $\omega$ is much slower than the spin relaxation rate $\tau_s^{-1}$, typically of the order of ${\rm ps}^{-1} = {\rm THz}$ in HMs.
Thus, the contribution to the reactance mainly comes from the effect of magnetization dynamics appearing in $G_\omega$.

\begin{figure}[tbp]
    \centering
    \includegraphics[width=7.5cm]{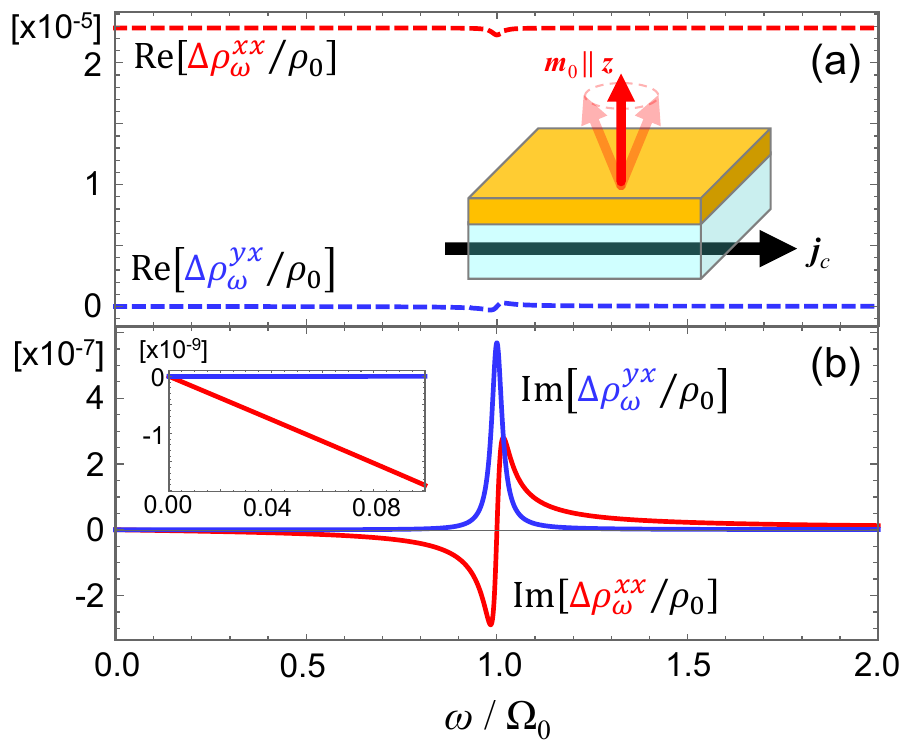}
    \caption{Frequency dependences of the (a) Real and (b) Imaginary parts of the longitudinal $(\Delta \rho_\omega^{xx})$ and transverse $(\Delta\rho_\omega^{yx})$ impedances,
    with out-of plane $\boldsymbol{m}_0 = \hat{\boldsymbol{z}}$.
    Inset of (a) shows schematic of the system, and
    inset of (b) shows ${\rm Im}[\Delta\rho_\omega^{xx,yx}/\rho_0]$ at low frequency.
    Parameters are taken as
    $\rho_0 = 1\;{\rm k\Omega} \; {\rm nm}$,
    $\theta_{\rm SH} = 0.01$,
    $\lambda_s = 1 \; {\rm nm}$,
    $\mu_0 M_s = 0.1 \; {\rm T}$,
    $\Omega_0 = 1 \; {\rm GHz}$,
    $\alpha = 0.01$,
    $d_H = d_F = 10 \; {\rm nm}$,
    and $(g_{\rm r}, g_{\rm i}) = (2,0)\times 10^{14} e^{-1} {\rm \Omega^{-1} m^{-2}}$.
    }
    \label{fig:fig2}
\end{figure}

Around the FMR frequency $\Omega_0$,
the impedance $\Delta\rho_\omega$ acquires a resonant behavior from $D_\omega$,
as pointed out in Refs.~\onlinecite{reiss2021theory} and \onlinecite{lotze2014spin}.
On the other hand,
for the off-resonant $\omega \ll \Omega_0$,
the reactance ${\rm Im}[\Delta\rho_\omega]$ becomes almost linear in $\omega$,
i.e., ${\rm Im}[\Delta\rho_\omega] = i\omega l + O(\omega^2)$,
because all the imaginary factors in Eq.~(\ref{eq:general-SMI}) appear in the form of $i\omega$
due to the Fourier transform of $\partial_t {\bm m}$.
Here, the tensor $l$ plays a role of an inductance tensor regularized by the system size.
For instance,
for the film with the length $d_x$ and the width $d_y$,
the longitudinal inductance becomes $L^{xx} = (d_x/d_y d_H) l^{xx}$.
Contrary to the classical solenoid inductors,
it is inversely proportional to the cross section $d_y d_H$ of the channel,
which is the behavior commonly seen in the preceding studies of emergent inductors \cite{nagaosa2019emergent}.
In particular, the resonance and the low-frequency behavior of ${\rm Im [\Delta\rho_\omega]}$ discussed above
are similar to those of the emergent inductance arising from the SOC in noncentrosymmetric electron systems,
such as Rashba and topological interfaces \cite{yamane2022theory,araki2023emergence}.
The spectrum of $\Delta\rho_\omega$ evaluated numerically from Eq.~(\ref{eq:general-SMI})
shown in Fig.~\ref{fig:fig2},
with $\boldsymbol{m}_0 = \hat{\bm z}$ and $g_{\rm i} =0$,
clearly shows these features.

\begin{figure}[tbp]
    \centering
    \includegraphics[width=7.5cm]{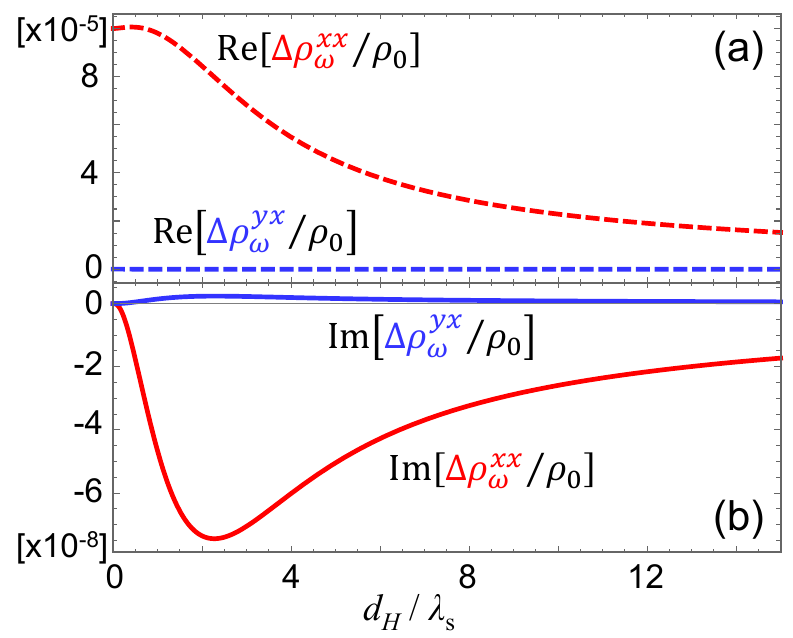}
    \caption{Dependences on the HM film thickness $d_H$
    for the (a) real and (b) imaginary parts of the longitudinal $(\Delta \rho_\omega^{xx})$ and transverse $(\Delta\rho_\omega^{yx})$ impedances.
    All the parameters except for $d_H$ are taken same as those employed in Fig.~\ref{fig:fig2},
    with the frequency fixed at $\omega = 0.7 \Omega_0$.
    }
    \label{fig:fig3}
\end{figure}

Dependence on the film thicknesses $d_H$ and $d_F$ is also important in the SHE-induced phenomena.
The $d_H$-dependence comes from the imbalance of spin accumulation in the film thickness direction,
and follows ${\rm Im}[\Delta\rho_\omega^{xx}] \propto \frac{\lambda_s}{d_H} \tanh^2 \frac{d_H}{2\lambda_s}$,
as shown in Fig.~\ref{fig:fig3}.
It shows a maximum at $d_H \sim 2\lambda_s$ and decays by $d_H^{-1}$ for $d_H \gg \lambda_s$,
similarly to the SMR in the DC limit.
Moreover, the reactance also shows $d_F$-dependence as ${\rm Im}[\Delta\rho_\omega^{xx}] \propto d_F^{-1}$,
which is not the case in the SMR.
(Note that the coherence of magnetization dynamics within the FM film is assumed in the present analysis.)
This is because the efficiency of the interfacial STT $\boldsymbol{\tau}^{\rm int}$,
which governs the magnetization dynamics essential for the reactance, is suppressed by the volume of the FM film.
\textcolor{black}{
    The film-thickness dependences discussed here will play an important role in distinguishing the present SHE-induced effect from other irrelevant effects in circuits experimentally, which we shall discuss in the last part of this paper.
}

\begin{figure}[tbp]
    \centering
    \includegraphics[width=7.5cm]{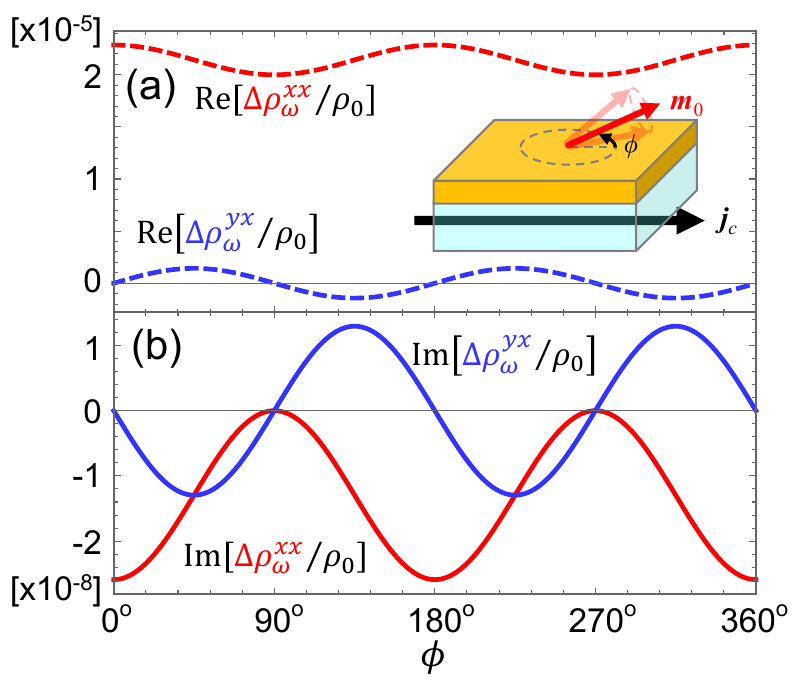}
    \caption{Dependences on the in-plane magnetic field direction $\boldsymbol{m}_0 = (\cos\phi, \sin\phi, 0)$
    for the (a) real and (b) imaginary parts of the longitudinal $(\Delta \rho^{xx}_\omega)$ and transverse $(\Delta\rho^{yx}_\omega)$ impedances.
    All the parameters except for $\boldsymbol{m}_0$ are taken same as those employed in Fig.~\ref{fig:fig2},
    with the frequency fixed at $\omega = 0.7 \Omega_0$.
    Inset of (a) shows schematic of the system.
    }
    \label{fig:fig4}
\end{figure}

The impedance $\Delta\rho_\omega$ is governed also by the magnetic properties of the FM:
it depends not only on the equilibrium magnetization direction $\boldsymbol{m}_0$,
but also on the magnetization dynamics via $D_\omega$.
Generally, $D_\omega$ is governed not only by the external magnetic field but also by the intrinsic magnetic properties,
such as magnetic anisotropy, demagnetization field, etc.
\textcolor{black}{
    If the demagnetization or magnetic anisotropy dominates over the external field,
    the FM will develop multiple domains to minimize the magnetostatic energy.
    In such cases, the dynamics of the domain walls may cause an additional contribution to the emergent reactance \cite{nagaosa2019emergent,ieda2021intrinsic},
    and hence the contribution from the magnetization dynamics considered here cannot be uniquely extracted.
    To avoid such a complexity, we here focus on the case where the external field $\boldsymbol{B}_{\rm ext}$ is dominant,
    so that the magnetic moments in the FM is uniformly aligned as $\boldsymbol{m}_0 \parallel \boldsymbol{B}_{\rm ext}$.
    Here, the magnetization dynamics is given by
}
$D_\omega = [i\omega -(\Omega_0 + i\alpha\omega)R_{\boldsymbol{m}_0}]^{-1}$
with $\Omega_0 \approx \gamma |\boldsymbol{B}_{\rm ext}|$.
In this limit,
the field-direction dependence of $\Delta\rho_\omega$ becomes identical to that of the SMR,
\begin{align}
    \Delta\rho_\omega^{xx} &= \Delta\rho^{(0)} + \Delta\rho_\omega^{(1)} [1-(m_0^y)^2] \label{eq:field-xx} \\
    \Delta\rho_\omega^{yx} &= \Delta\rho_\omega^{(1)} m_0^x m_0^y + \Delta\rho_\omega^{(2)} m_0^z \label{eq:field-yx},
\end{align}
where $\Delta\rho_\omega^{(1,2)}$ exhibit imaginary parts at finite $\omega$.
In particular, for the in-plane field direction $\boldsymbol{m}_0 = (\cos\phi, \sin\phi, 0)$,
both ${\rm Re}[\Delta\rho_\omega^{xx}]$ and ${\rm Im}[\Delta\rho_\omega^{xx}]$ show the $\cos 2\phi$ structure,
as demonstrated in Fig.~\ref{fig:fig4}.
We should also note that the reactance acquires a transverse component ${\rm Im}[\Delta\rho_\omega^{yx}]$ depending on $\boldsymbol{m}_0$.
It consists of the planar Hall component $(\propto m_0^x m_0^y)$ and the anomalous Hall component $(\propto m_0^z)$,
as in the case of DC SMR.
The anomalous Hall component in the reactance
is pointed out in the SOC-induced emergent inductance as well \cite{yamane2022theory,araki2023emergence},
which would serve as an important feature in identifying the symmetry structure of the electron system hosting the SOC.

\begin{figure}[tbp]
    \centering
    \includegraphics[width=7cm]{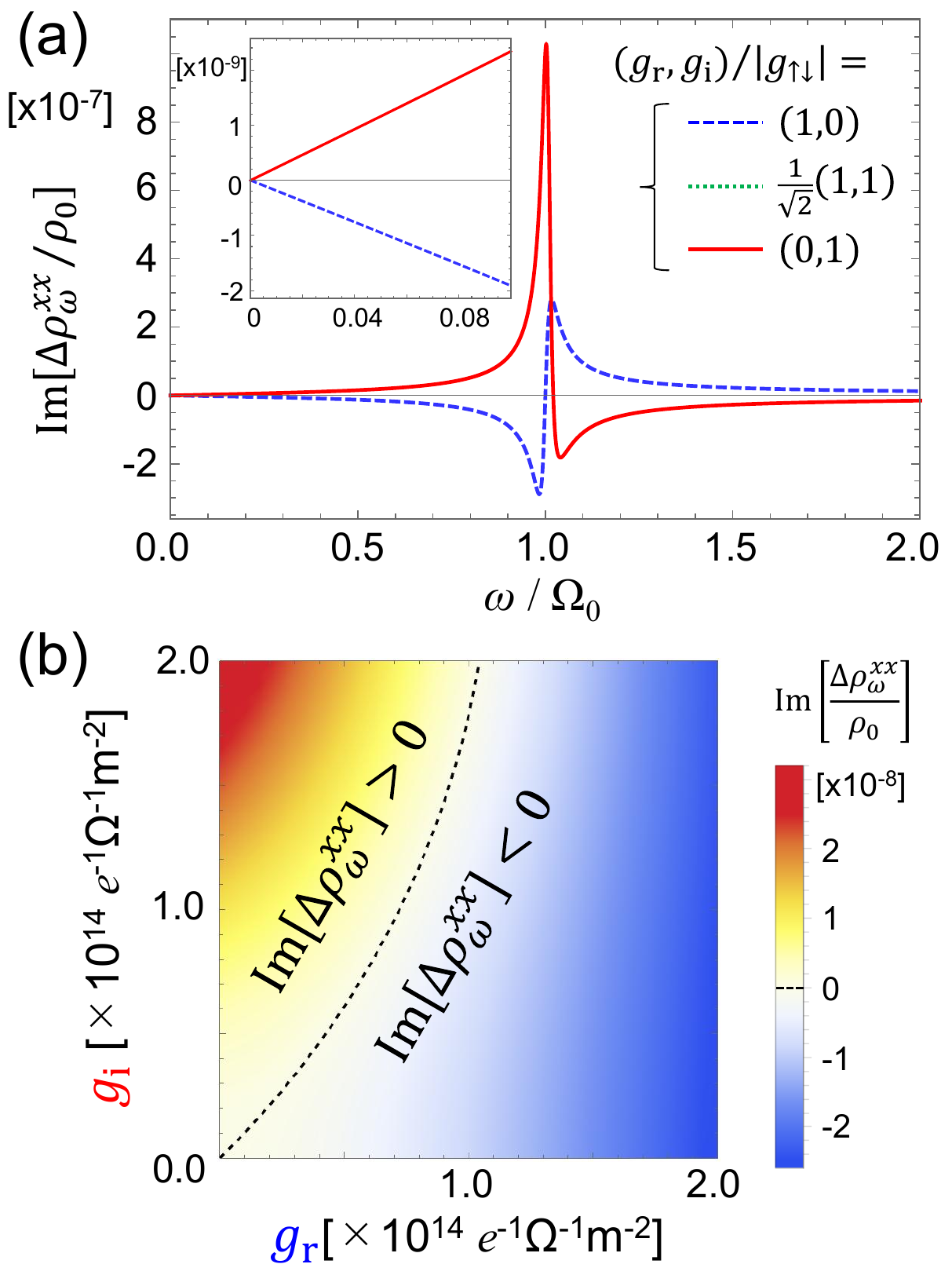}
    \caption{(a) Frequency dependence of the longitudinal reactance ${\rm Im}[\Delta\rho_\omega^{xx}]$ for three different ratios of the spin mixing conductance $(g_{\rm r}, g_{\rm i})$,
    with $|g_{\uparrow\downarrow}| = 2\times 10^{14} e^{-1} {\rm \Omega^{-1} m^{-2}}$ fixed.
    Inset shows their behaviors at low frequency.
    (b) Color map of ${\rm Im}[\Delta\rho_\omega^{xx}]$ evaluated at $\omega = 0.7 \Omega_0$,
    with $(g_{\rm r}, g_{\rm i})$ varied.
    All the parameters except for $g_{\uparrow\downarrow} = g_{\rm r} + i g_{\rm i}$ are taken same as those employed in Fig.~\ref{fig:fig2}.
    }
    \label{fig:fig5}
\end{figure}

The reactance process consists of the combination of the spin injection $\boldsymbol{j}_s^{\rm inj}$ and spin pumping $\boldsymbol{j}_s^{\rm pump}$,
both of which are governed by the interfacial spin mixing conductance $g_{\uparrow\downarrow}$.
Thus, ${\rm Im}[\Delta\rho_\omega]$ becomes at least of $O(g_{\uparrow\downarrow}^2)$.
This is in a clear contrast with the case of SMR,
which is of $O(g_{\uparrow\downarrow})$ because it emerges solely from $\boldsymbol{j}_s^{\rm inj}$.
In particular, for $g_{\uparrow\downarrow} \ll {\sigma_0}/{2e\lambda_s}$,
the leading-order contributions of $O(g_{\uparrow\downarrow}^2)$ to the components ${\rm Im}[\Delta\rho_\omega^{(1,2)}]$ in Eqs.~(\ref{eq:field-xx}) and (\ref{eq:field-yx}) read (see Supplemental material for the derivation process),
\begin{align}
    {\rm Im}
    \begin{pmatrix}
        \Delta\rho_\omega^{(1)} \\ \Delta\rho_\omega^{(2)}
    \end{pmatrix}
    &\approx \frac{\omega}{\Omega_0} \frac{\gamma}{M_s d_F d_H} \left[ \frac{\theta_{\rm SH} \hbar \lambda_s}{\sigma_0} \tanh \frac{d_H}{2\lambda_s} \right]^2
    \begin{pmatrix}
        g_{\rm i}^2 - g_{\rm r}^2 \\ 2g_{\rm r} g_{\rm i}
    \end{pmatrix}
    .
    \label{eq:gi-gr}
\end{align}
This form indicates that the sign of the longitudinal reactance,
${\rm Im}[\Delta\rho_\omega^{xx}] = {\rm Im}[\Delta\rho_\omega^{(1)}] [1-(m_0^y)^2]$,
is determined by the relative magnitude between $g_{\rm r}$ and $g_{\rm i}$:
the reactance becomes negative if the damping-like component (corresponding to $g_{\rm r}$) is dominant,
and becomes positive if the field-like component (corresponding to $g_{\rm i}$) is dominant,
as can be seen in Fig.~\ref{fig:fig5}(a).
\textcolor{black}{
    This tendency remains unchanged by the interfacial spin memory loss,
    because the spin memory loss does not change the ratio between $g_{\rm r}$ and $g_{\rm i}$ but only renormalizes $|g_{\uparrow\downarrow}|$ \cite{rojas2014spin}.
}
\textcolor{black}{
    Although ${\rm Im}[\Delta\rho_\omega]$ deviates from the $O(g_{\uparrow\downarrow}^2)$-like behavior for $|g_{\uparrow\downarrow}| \gtrsim {\sigma_0}/{2e\lambda_s}$ (see Supplemental Material),
    the sign-change behavior of ${\rm Im}[\Delta\rho_\omega^{xx}]$ depending on the ratio between $g_{\rm r}$ and $g_{\rm i}$ is still present,
    as shown in Fig.~\ref{fig:fig5}(b) from the full numerical evaluation of Eq.~(\ref{eq:general-SMI}).
}
\textcolor{black}{
    While $g_{\rm i} < g_{\rm r}$ is supposed in commonly used HM/FM heterostructures such as Pt/YIG,
    $g_{\rm i} > g_{\rm r}$ is also experimentally reported in several combinations of HM/FM (e.g., W/EuO) \cite{rosenberger2021quantifying,dubowik2020non},
    in which we expect ${\rm Im}[\Delta\rho_\omega^{xx}] >0$ to be potentially accessible. 
}

The sign change in the reactance seen above can be compared with those found in the preceding theories of emergent inductances.
In particular, it is similar to the SOC-induced emergent inductance in noncentrosymmetric electronic systems,
in which the field-like and damping-like components of the SOT and SMF give contributions of the opposite signs to the emergent inductance \cite{yamane2022theory}.
With the nonadiabaticity parameter $\beta$ characterizing the relative magnitude between them,
the emergent inductance becomes proportional to $1-\beta^2$,
which is compatible to $g_{\rm i}^2 - g_{\rm r}^2$ found in Eq.~(\ref{eq:gi-gr}) here.
Also in the emergent inductance arising from magnetic texture dynamics,
the competition between the adiabatic and nonadiabatic STT and SMF is predicted to yield the sign change in the inductance \cite{ieda2021intrinsic}, 
which are identified as the intraband and interband effects in terms of the microscopic band structure of electrons \cite{kurebayashi2021electromagnetic,anan2025emergent}.
All the sign change behaviors in the reactance (inductance) seen so far can be traced back to the competition of two excitation modes that are conjugate to each other:
two perpendicular magnetization components (e.g., $m_x$ vs $m_y$) for the dynamics of uniform magnetization including the present study,
and the sliding motion vs the phase rotation for the dynamics of magnetic spirals and domain walls.


With the material parameters for typical HM/FM heterostructures,
the reactance for the sample size $d_x = 1\ {\rm mm}, d_y = 1\ {\rm \mu m}, d_H=d_F = 10\ {\rm nm}$
is estimated to be of the inductance $L^{xx} \approx 10\ {\rm pH}$ (see Supplemental Material for detailed calculation).
\textcolor{black}{
    While this inductance value appears to be small compared with those of conventional inductors (of the orders of nH-$\mu$H),
    it yields a non-negligible change in the resonance spectrum of an LCR circuit,
    which we expect to be detected by frequency-resolved measurements.
}
In \textcolor{black}{such} measurements, one needs to
separate other possible contributions to reactance that are irrelevant to magnetization dynamics,
such as \textcolor{black}{the classical inductance arising from the Oersted field,}
the resistivity modulation by Joule heating \cite{furuta2024reconsidering},
and parasitic capacities in circuits \cite{furuta2023energetic,choi2024questioning}.
\textcolor{black}{
    To distinguish them experimentally,
    their dependences on the system parameters,
    especially the dependences on film thicknesses $(d_H,d_F)$,
    will play important roles.
    The classical inductance contribution is expected to become almost independent of $d_H$,
    because the Oersted field strength depends not on the current density but on the net current in the HM film.
    The Joule heating contribution will become almost independent of $d_F$,
    because it is governed by the resistivity of the HM film and the system's heat capacity usually dominated by large substrates.
    Furthermore,
    the parasitic capacity occurs mostly in the circuit configuration outside the HM/FM sample,
    and hence it will not depend on either $d_H$ or $d_F$.
    Those parameter dependences are distinct from the behavior of the SHE-induced reactance studied throughout this work,
    and hence we expect their contributions to be systematically ruled out in experimental measurements.
}

\textcolor{black}{
    After the separation of parasitic effects discussed above, 
    the reactance measurement would be useful to understand the nature of SHE-induced spin dynamics in the FM film.
    Besides the uniform FMR mode dominating in the thin-film regime,
    the magnetostatic standing spin waves participate in the spin injection and pumping processes
    significantly in a thick FM film $(d_F \gtrsim 1 {\rm \mu m})$ \cite{sandweg2010enhancement,chen2019incoherent}.
    In this regime, the frequencies of those magnetostatic modes are well separated from that of the uniform FMR,
    which potentially contribute to the reactance as well.
    Moreover, in microstructures such as the vertical meander waveguides and lateral arrays,
    some magnon modes of finite wavelengths can emerge even below the frequency of the uniform FMR \cite{gubbiotti2021magnonic,gubbiotti2021magnonic2,sadovnikov2018neuromorphic}.
    Since such modes reside at lateral length scales of 1-100 $\mu{\rm m}$,
    they can be also excited by the spin Hall effect in nanoscale devices,
    which may give an indispensable contribution to the reactance.
    Solid understanding on the reactance emerging from those nonuniform modes is left for future detailed analysis.
}

\begin{acknowledgments}
The authors thank Takumi Funato, Takashi Kikkawa, Maki Umeda, and Yuta Yamane for fruitful discussions.
This work was supported by Japan Society for the Promotion of Science KAKENHI
(JP22K03538, JP23K17882, JP24H00409, JP25K01298).
\end{acknowledgments}

\bibliography{SMI-manuscript.bib}

~
\clearpage

\appendix

\appendix
\renewcommand{\thefigure}{A\arabic{figure}}
\setcounter{figure}{0} 

\section{Detail of derivation}
In this part, we review the derivation process of the impedance tensor $\Delta\rho_\omega$ in Eq.~(\ref{eq:general-SMI}),
following the finite-frequency formalism established in Refs.~\onlinecite{reiss2021theory} and \onlinecite{chiba2014current}.
We here employ the tensor forms of $90^\circ$-rotations,
\begin{align}
    R_{\hat{\boldsymbol{z}}} = \begin{pmatrix}
        0 & -1 & 0 \\
        1 & 0 & 0 \\
        0 & 0 & 0
    \end{pmatrix}
    , \quad
    R_{\boldsymbol{m}_0} = \begin{pmatrix}
        0 & -m_0^z & m_0^y \\
        m_0^z & 0 & -m_0^x \\
        -m_0^y & m_0^x & 0
    \end{pmatrix}
\end{align}
in the Cartesian coordinate.

By substituting $\boldsymbol{j}_s^z$ in Eq.~(\ref{eq:response-js}) to Eq.~(\ref{eq:spin-diffusion}),
we obtain the spin diffusion equation,
\begin{align}
    \left[ i\omega +\tau_s^{-1} - D_s \partial_z^2 \right] \boldsymbol{\mu}_{s\omega}(z) &= 0,
\end{align}
with the frequency representation.
The general solution for this equation is given as,
\begin{align}
    \boldsymbol{\mu}_{s\omega}(z) &= \boldsymbol{\mu}_{s\omega}^{(1)} \cosh\frac{z}{\lambda_\omega} - \boldsymbol{\mu}_{s\omega}^{(2)} \sinh\frac{z}{\lambda_\omega}, \label{eq:spin-solution}
\end{align}
with $\lambda_\omega = \lambda_s / \sqrt{1+i\omega\tau_s}$ and $\lambda_s = \sqrt{D\tau_s}$.
On the other hand, the general solution for the LLG equation [Eq.~(\ref{eq:LLG})] linearized for $\boldsymbol{u}(t) = \boldsymbol{m}(t) - \boldsymbol{m}_0$ becomes,
\begin{align}
    \boldsymbol{u}_\omega = D_\omega \boldsymbol{\tau}^{\rm int}_\omega, \label{eq:u-solution}
\end{align}
with the Green's function $D_\omega$.

The solutions of Eqs.~(\ref{eq:spin-solution}) and (\ref{eq:u-solution}) are uniquely determined by the boundary conditions imposed at the interfaces.
At the vacuum interface $z = d_H$,
the boundary condition $\boldsymbol{j}_s^z(z=d_H) = 0$ gives a restriction for $\boldsymbol{\mu}_{s\omega}$ [Eq.~(\ref{eq:spin-solution})],
\begin{align}
    -\frac{\sigma_0}{2e\lambda_\omega} \left[ \boldsymbol{\mu}_{s\omega}^{(1)} \sinh\frac{d_H}{\lambda_\omega} - \boldsymbol{\mu}_{s\omega}^{(2)} \cosh\frac{d_H}{\lambda_\omega} \right] -\theta_{\rm SH} \sigma_0 R_{\hat{\boldsymbol{z}}}\boldsymbol{E}_\omega = 0. \label{eq:boundary-dH}
\end{align}
On the other hand, at the HM-FN interface $z = 0$,
$\boldsymbol{j}_s^{\rm inj}$ and $\boldsymbol{j}_s^{\rm pump}$ in Eqs.~(\ref{eq:j-inj}) and (\ref{eq:j-pump}) become,
\begin{align}
    \boldsymbol{j}_{s\omega}^{\rm inj} &= -G_s \boldsymbol{\mu}_{s\omega}(z=0) = -G_s \boldsymbol{\mu}_{s\omega}^{(1)} \\
    \boldsymbol{j}_{s\omega}^{\rm pump} &= \hbar G_s R_{\boldsymbol{m}_0} (i\omega \boldsymbol{u}_\omega) \\
    &= i\hbar\omega G_s R_{\boldsymbol{m}_0} D_\omega \left[ \frac{\gamma}{M_s d_F} \frac{\hbar}{2e} \boldsymbol{j}_{s\omega}^{\rm int} \right]
\end{align}
by using Eqs.~(\ref{eq:spin-solution}), (\ref{eq:u-solution}), and $\boldsymbol{\tau}^{\rm int}$ introduced in Eq.~(\ref{eq:t-int}).
Thus,  we obtin a relation between the coeffcient $\boldsymbol{\mu}_{s\omega}^{(1)}$ and $\boldsymbol{j}_{s}^{\rm int} = \boldsymbol{j}_s^{\rm inj} + \boldsymbol{j}_s^{\rm pump}$ as,
\begin{align}
    \left[ 1- i\omega \frac{\hbar^2}{2e}  \frac{\gamma}{M_s d_F} G_s R_{\boldsymbol{m}_0} D_\omega \right] \boldsymbol{j}_{s\omega}^{\rm int} &= -G_s \boldsymbol{\mu}_{s\omega}^{(1)},
\end{align}
yielding $\boldsymbol{j}_{s\omega}^{\rm int} = -G_\omega \boldsymbol{\mu}_{s\omega}^{(1)}$ with $G_\omega$ defined by Eq.~(\ref{eq:G-omega}).
By using this $\boldsymbol{j}_{s\omega}^{\rm int}$,
the boundary condition $\boldsymbol{j}_{s}^{\rm int} = \boldsymbol{j}_{s}^z(z=0)$ can be reduced as,
\begin{align}
    -G_\omega \boldsymbol{\mu}_{s\omega}^{(1)} &= \frac{\sigma_0}{2e\lambda_\omega} \boldsymbol{\mu}_{s\omega}^{(2)}  -\theta_{\rm SH} \sigma_0 R_{\hat{\boldsymbol{z}}} \boldsymbol{E}_\omega. \label{eq:boundary-0}
\end{align}
After all, the forms of $\boldsymbol{\mu}_{s\omega}^{(1,2)}$ are uniquely determined by
the two equations Eqs.~(\ref{eq:boundary-dH}) and (\ref{eq:boundary-0}) ,
\begin{align}
    \boldsymbol{\mu}_{s\omega}^{(1)} &= 2e\lambda_\omega \theta_{\rm SH} \frac{\coth\tfrac{d_H}{\lambda_\omega} - {\rm csch}\tfrac{d_H}{\lambda_\omega}}{1 + \frac{2e\lambda_\omega}{\sigma_0} \coth\tfrac{d_H}{\lambda_\omega} G_\omega} R_{\hat{\boldsymbol{z}}} \boldsymbol{E}_\omega, \\
    \boldsymbol{\mu}_{s\omega}^{(2)} &= 2e\lambda_\omega \theta_{\rm SH} \frac{1 + \frac{2e\lambda_\omega}{\sigma_0} \mathrm{csch}\tfrac{d_H}{\lambda_\omega} \ G_\omega}{1 + \frac{2e\lambda_\omega}{\sigma_0} \coth\tfrac{d_H}{\lambda_\omega} G_\omega} R_{\hat{\boldsymbol{z}}} \boldsymbol{E}_\omega.
\end{align}

With $\boldsymbol{\mu}_{s\omega}^{(1,2)}$ obtained above,
we can now evaluate the contribution to the impedance.
By substituting the solution Eq.~(\ref{eq:spin-solution}) to Eq.~(\ref{eq:response-jc}),
we obtain the electric current distribution $\boldsymbol{j}_{c\omega}(z) = \sigma_0 \boldsymbol{E}_\omega + \Delta\boldsymbol{j}_{c\omega}(z)$, with
\begin{align}
    \Delta\boldsymbol{j}_{c\omega}(z) &= \frac{\theta_{\rm SH} \sigma_0}{2e \lambda_\omega} R_{\hat{\boldsymbol{z}}} \left[ \boldsymbol{\mu}_{s\omega}^{(1)} \sinh\frac{z}{\lambda_\omega} - \boldsymbol{\mu}_{s\omega}^{(2)} \cosh\frac{z}{\lambda_\omega} \right],
\end{align}
and hence its average over the film thickness becomes
\begin{align}
    \Delta\boldsymbol{j}_{c\omega}^{\rm ave} &= \frac{\theta_{\mathrm{SH}} \sigma_0}{2e d_H} R_{\hat{\boldsymbol{z}}} \left[ \boldsymbol{\mu}_{s\omega}^{(1)} \left( \cosh\frac{d_H}{\lambda_\omega}-1 \right) - \boldsymbol{\mu}_{s\omega}^{(2)} \sinh\frac{d_H}{\lambda_\omega} \right].
\end{align}
By substituting $\boldsymbol{\mu}_{s\omega}^{(1,2)}$ obtained above,
we obtain the form of the SHE-induced conductivity $\Delta\sigma_\omega$ satisfying $\Delta\boldsymbol{j}_{c\omega}^{\rm ave} = \Delta\sigma_\omega E_\omega$,
from which we can derive the impedance tensor $\Delta\rho_\omega = -\sigma_0^{-2} \Delta\sigma_\omega$ as shown in Eq.~(\ref{eq:general-SMI}).

In the low-frequency off-resonant regime $\omega \ll \Omega_0$,
$D_\omega$ and $\lambda_\omega$ reduce to $D_0 = D_{\omega=0}$ and $\lambda_s = \lambda_{\omega=0}$, respectively.
Thus, $\Delta\rho_\omega$ can be expanded by $\omega$ as
\begin{align}
    \Delta\rho_\omega = \Delta\rho_0 + i\omega l + O(\omega^2),
\end{align}
\begin{widetext}
\noindent
where the ``inductance'' tensor $l$ for the reactance ${\rm Im}[\Delta\rho_\omega] = \omega l + O(\omega^2)$ is given as
\begin{align}
    l &= -\rho_0 \theta_{\rm SH}^2 \frac{\hbar^2}{2e} \frac{\gamma}{M_s d_F} \frac{2e\lambda_s}{\sigma_0} \frac{\lambda_s}{d_H} \tanh^2\frac{d_H}{2\lambda_s} R_{\hat{\boldsymbol{z}}} \left[\frac{G_s}{\boldsymbol{1} + \frac{2e\lambda_s}{\sigma_0}G_s \coth\frac{d_H}{\lambda_s}}\right] R_{\boldsymbol{m}_0} D_0 \left[\frac{G_s}{\boldsymbol{1} + \frac{2e\lambda_s}{\sigma_0}G_s \coth\frac{d_H}{\lambda_s}} \right] R_{\hat{\boldsymbol{z}}}. \label{eq:l-general}
\end{align}
\end{widetext}
This is the most general expression for $l$,
without any approximation on the structure of $D_0$ and $G_s$.
If the spin mixing conductance is small enough $(|g_{\uparrow\downarrow}| \ll \tfrac{\sigma_0}{2e\lambda_s})$,
The denominators $1 + \frac{2e\lambda_s}{\sigma_0}G_s \coth\frac{d_H}{\lambda_s}$ in Eq.~(\ref{eq:l-general}) reduce to $1$,
and $l$ becomes of $O(g_{\uparrow\downarrow}^2)$.

If the external magnetic field is dominant over the demagnetization field,
the equilibrium magnetization $\boldsymbol{m}_0$ aligns along the external field $\boldsymbol{B}_{\rm ext}$,
and the Green's function $D_\omega$ becomes,
\begin{align}
    D_\omega &= \left[ i\omega - \gamma (\boldsymbol{B}_{\rm ext}\times) - i\omega\alpha (\boldsymbol{m}_0 \times)  \right]^{-1} \\
    &= \left[ i\omega - (\Omega_0 + i\omega\alpha)R_{\boldsymbol{m}_0} \right]^{-1},
\end{align}
with $\Omega_0 = \gamma |\boldsymbol{B}_{\rm ext}|$.
In this case, both $D_\omega$ and $G_s$ (and hence $G_\omega$ as well) are described as powers of $R_{\boldsymbol{m}_0}$,
and thus Eq.~(\ref{eq:general-SMI}) can be reduced as,
\begin{align}
    \frac{\Delta\rho_\omega}{\rho_0} &= -\theta_{\rm SH}^2 \frac{\lambda_\omega}{d_H} R_{\hat{\boldsymbol{z}}} \left[ 2\eta_\omega(\tfrac{d_H}{2}) + X_\omega R_{\boldsymbol{m}_0}^2 + Y_\omega R_{\boldsymbol{m}_0} \right] R_{\hat{\boldsymbol{z}}}, \label{eq:SMI-isotropic}
\end{align}
with the numerical factors $X_\omega$ and $Y_\omega$
(Note that $R_{\boldsymbol{m}_0}$ satisfies $R_{\boldsymbol{m}_0}^3 = -R_{\boldsymbol{m}_0}$).
The first, second, and third terms in Eq.~(\ref{eq:SMI-isotropic}) give the components $\Delta\rho^{(0)}$, $\Delta\rho_\omega^{(1)}$, and $\Delta\rho_\omega^{(2)}$ in Eqs.~(\ref{eq:field-xx})(\ref{eq:field-yx}), respectively.

By using $D_0 = (1/\Omega_0) R_{\boldsymbol{m}_0}$ in the zero-frequency limit,
the inductance tensor $l$ in Eq.~(\ref{eq:l-general}) with $|g_{\uparrow\downarrow}| \ll \tfrac{\sigma_0}{2e\lambda_s}$ further reduces as,
\begin{align}
    l &= \rho_0 \theta_{\rm SH}^2 \frac{\hbar^2}{2e} \frac{\gamma}{M_s \Omega_0 d_F} \frac{2e\lambda_s}{\sigma_0} \frac{\lambda_s}{d_H} \tanh^2\frac{d_H}{2\lambda_s} R_{\hat{\boldsymbol{z}}} G_s^2 R_{\hat{\boldsymbol{z}}} \nonumber \\
    &= \frac{\hbar^2 \theta_{\rm SH}^2 \gamma \lambda_s^2}{M_s \Omega_0 d_F d_H \sigma_0^2} \tanh^2\frac{d_H}{2\lambda_s} \\
    & \hspace{20pt} \times R_{\hat{\boldsymbol{z}}} \left[ (g_{\rm i}^2 - g_{\rm r}^2)R_{\boldsymbol{m}_0}^2 - 2g_{\rm r}g_{\rm i} R_{\boldsymbol{m}_0} \right] R_{\hat{\boldsymbol{z}}} \nonumber
\end{align}
\textcolor{black}{up to $O(g_{\uparrow\downarrow}^2)$,}
which gives Eq.~(\ref{eq:gi-gr}) in the main text.

\begin{figure}[tbp]
    \centering
    \includegraphics[width=8cm]{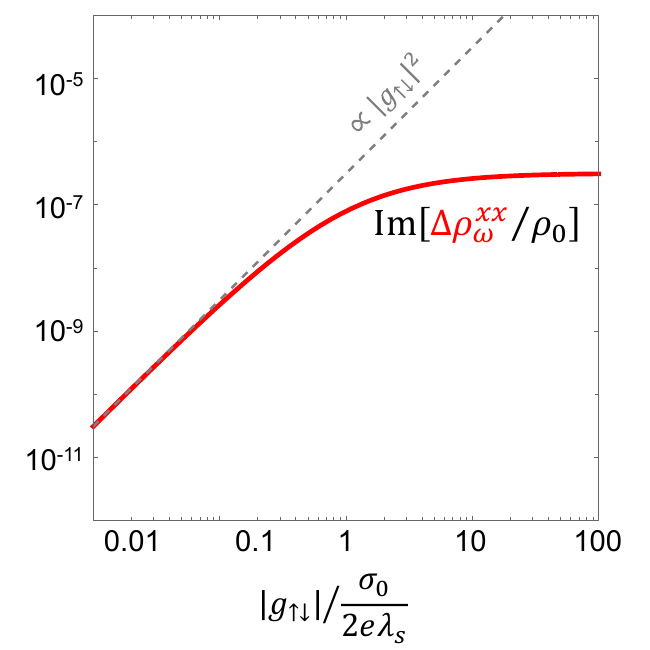}
    \caption{
        \textcolor{black}{
            Log-Log plot for the dependence of ${\rm Im}[\Delta\rho_\omega^{xx}]$ on $|g_{\uparrow\downarrow}|/\tfrac{\sigma_0}{2e\lambda_s}$ to show the deviation from the perturbatively obtained $O(g_{\uparrow\downarrow}^2)$-like behavior.
        }
    }
    \label{fig:rho-g}
\end{figure}

\textcolor{black}{
    For $|g_{\uparrow\downarrow}| \gtrsim \tfrac{\sigma_0}{2e\lambda_s}$,
    the value of ${\rm Im}[\Delta\rho_\omega]$ (or $l$) deviates from the $O(g_{\uparrow\downarrow}^2)$-like behavior in Eq.~(12) given by the perturbative expansion.
    Figure \ref{fig:rho-g} demonstrates this deviation as a log-log plot for ${\rm Im}[\Delta\rho_\omega^{xx}]$ under $\boldsymbol{m}_0 \parallel x$,
    where the red curve shows the behavior of ${\rm Im}[\Delta\rho_\omega^{xx}]$ fully evaluated by Eq.~(8),
    and the gray dashed line corresponds to the perturbatively obtained ${\rm Im}[\Delta\rho_\omega^{xx}]$ given by Eq.~(12).
    We here fix $\omega = 0.7 \Omega_0$, and take all the other parameters same as those employed in Fig.~2.
    We see from this plot that ${\rm Im}[\Delta\rho_\omega^{xx}]$ deviates from the $O(g_{\uparrow\downarrow}^2)$-like behavior for $|g_{\uparrow\downarrow}|/\tfrac{\sigma_0}{2e\lambda_s} \gtrsim 1$,
    and saturates to a finite value for $|g_{\uparrow\downarrow}|/\tfrac{\sigma_0}{2e\lambda_s} \gtrsim 10$.
}

\section{Order estimation of reactance}
In the numerical estimation of the magnitude of reactance, we use the set of material parameters as follows:
\begin{itemize}
    \item HM film:
    \begin{itemize}
        \item Resistivity $\rho_0 = 1 \;{\rm k\Omega}\;{\rm nm}$
        \item Spin Hall angle $\theta_{\rm SH} = 0.01$
        \item Spin diffusion length $\lambda_s = 1 \;{\rm nm}$
    \end{itemize}
    \item FM film:
    \begin{itemize}
        \item Saturation magnetization $\mu_0 M_s = 0.1 \; {\rm T}$
        \item FMR frequency $\Omega_0 = 1 \;{\rm GHz}$
        \item Gilbert damping constant $\alpha = 0.01$
    \end{itemize}
    \item Interface:
    \begin{itemize}
        \item Spin mixing conductance \\ $g_{\uparrow\downarrow} = g_{\rm r} + i g_{\rm i} = (2+0i) \times 10^{14} e^{-1} {\rm \Omega}^{-1} {\rm m}^{-2}$ .
    \end{itemize}
\end{itemize}
The orders of these parameters are comparable to those used for the preceding theories of SMR,
considering Pt for HM and ${\rm Y}_3 {\rm Fe}_5 {\rm O}_{12}$ (yttrium iron garnet, YIG) for FM \cite{chen2013theory,reiss2021theory}.

With $d_H = d_F = 10 \;{\rm nm}$,
the numerical calculation based on Eq.~(\ref{eq:general-SMI}) gives ${\rm Im}[\Delta\rho_\omega^{xx}/\rho_0] \approx 10^{-7}$ at $\omega = 0.9 \Omega_0$.
This value corresponds to the volume-normalized inductance,
$l^{xx} = {{\rm Im}[\Delta\rho_\omega^{xx}]}/{\omega} \approx 0.1 \; {\rm pH}\;{\rm nm}$.
Therefore, with the system size $d_x = 1\;{\rm mm}$ and $d_y = 1\;{\rm \mu m}$,
the order of the inductance becomes,
$L^{xx} = l^{xx} \times (d_x/d_y d_H) \approx 10 \;{\rm pH}$.

\vspace{10cm}

\end{document}